\documentclass{article}[12pt]
\usepackage{graphicx}
\usepackage{setspace}
 \DeclareGraphicsExtensions{.eps, .ps}

\def\pr{\mathop{\rm pr}\nolimits}

\def\cov{\mathop{\rm cov}\nolimits}

\def\GP{\mathop{\rm GP}\nolimits}

\def\indep{\mathrel{\rlap{$\perp$}\kern1.6pt\mathord{\perp}}}
\newbox\dashbox
\setbox\dashbox=\vbox to 6pt{\vfill\hbox to 0.4cm{\hrulefill}\vfill}

\def\H{{\cal H}}

\def\State{{\cal S}}
\def\given{\mathrel{|}}
\def\Given{\mathrel{\Big |}}
\def\bft{{\bf t}}

\mathchardef\Real="023C

\newif\ifignoretext \ignoretexttrue
\def\beginignoretext{\setbox0=\vbox\bgroup}
\def\endignoretext{\egroup \ifignoretext\relax\else\unvbox0 \fi}

\newbox\bigstrutbox
\setbox\bigstrutbox=\hbox{\vrule height 10.5pt depth3.5pt width 0pt}
\def\bigstrut{\relax\ifmmode\copy\bigstrutbox\else\unhcopy\bigstrutbox\fi}
\chardef\tie='176
\chardef\caret='136

\begin{document}
\title{Vital variables and survival processes}
\author{Walter Dempsey and Peter McCullagh\thanks{Department of Statistics, University of Chicago, 5734
University Ave, Chicago, Il 60637, U.S.A. E-mail:
pmcc@galton.uchicago.edu}}
\maketitle

\begin{abstract}\noindent
The focus of a survival study is partly on the distribution of survival times,
and partly on the health or quality of life of patients while they live.
Health varies over time, and survival is the most basic aspect of health,
so the two aspects are closely intertwined.
Depending on the nature of the study, a range of variables may be measured;
some constant in time, others not;
some regarded as responses, others as explanatory risk factors;
some directly and personally health-related, others less directly so.
This paper begins by classifying variables that may arise in such a setting, emphasizing
in particular, the mathematical distinction between vital and non-vital variables.
We examine also various types of probabilistic relationships that may exist among variables.
Independent evolution is an asymmetric relation, which is
intended to encapsulate the notion of one process driving the other;
$X$~is a driver of~$Y$ if $X$ evolves independently of the history of~$Y$.
This concept arises in several places in the study of survival processes.
\end{abstract}

\section{Introduction}
This paper is concerned with survival times and health monitoring,
meaning mathematical models for the distribution of survival times and health-related processes.
Health monitoring is interpreted broadly to include the measurement of any variable
that might be deemed relevant to individual health.
Obvious examples include
(i)~blood serum level,
(ii)~pulse rate,
(iii)~quality of life in a geriatric study, or
(iv)~physical dexterity or mental acuity.
Less obvious examples include
(v)~the value of the patient's retirement portfolio,
(vi)~the ambient temperature or ozone level as a measure of health threat,
or (vii)~the patient's schedule of medical appointments.
An intermediate example is (viii)~the type of care facility or retirement home.
The focus is on situations where mortality is appreciable,
so geriatric studies feature prominently in examples.

Mathematically speaking, a variable $Y$ is a function of time,
so that $Y_i(t)$ is the value for patient~$i$ at time $t\ge 0$,
measured relative to a suitable temporal origin, usually recruitment.
If $T_i > 0$ is the survival time for patient~$i$, the re-coded variable
\[
\tilde T_i(t) = \left\{ \begin{array}{rl} 1 &\quad t < T_i \cr 0 &\quad \hbox{otherwise}, 
	\end{array} \right.
\]
is a function of time, defined for all $t \ge 0$.
As the re-coded survival process illustrates, every variable $Y$ is
defined for all $t\ge 0$, including $t > T_i$.

In some contexts such as the retirement portfolio or the ozone pollution level,
the definition for $t > T_i$ poses no physical or conceptual difficulty;
the process evolves after death and values can be recorded with little additional inconvenience.
If the portfolio has been liquidated or dispersed, the value is zero.
In other instances where it is not feasible to measure the pulse rate or quality of life
without detecting that the patient is no longer alive,
it is necessary to include in the state space a special value indicating that
the patient is dead.
Death is presumed to be a fatal event, so that this special value is an absorbing state.
For example, $\tilde T_i(t) = 0$ is the absorbing state for survival.

\section{Vital and non-vital variables}
Let $Y_i \equiv (Y_i(s))_{s\ge 0}$ be the entire temporal trajectory of the variable~$Y$ for
patient~$i$, and let $T_i > 0$ be the survival time.
The variable $Y$ is said to be \emph{vital} if the single value $Y_i(t)$ suffices to determine
whether or not patient~$i$ is alive at time~$t$.
In other words, $Y$ is vital if, for each $t\ge 0$, the conditional probability satisfies
\begin{equation}\label{vital}
\pr(T_i > t \given Y) = \pr(T_i > t \given Y_i(t)) \in \{0, 1\}.
\end{equation}
The first part of the condition asserts
that the event $T_i > t$ is conditionally independent of $Y$ given $Y_i(t)$,
so the single value $Y_i(t)$ suffices to determine the conditional probability.
The second part says that the conditional probability is either zero or one,
i.e.,~no intermediate values are allowed.
Otherwise, if  
the present value alone is not sufficient,
or if the conditional probability lies in $(0,1)$ for some~$t$,
we say that $Y$ is \emph{non-vital}.

For survival studies, the distinction between vital and non-vital variables is fundamental.
It is immediately apparent that the re-coded survival time $\tilde T$ is vital.
In addition, if $Y$~is vital, and $Z$~is any other variable defined concurrently,
the pair $(Y,Z)$ is also vital.
Likewise, if $Z$~is real or vector-valued with no atom at zero, the product $\tilde T Z$ is vital.
Thus, a vital variable may contain components that are non-vital, irrelevant, or even trivial.

A vital variable that is constant post mortem is called purely vital.
It has one or more absorbing states, each representing death, but perhaps associated with different causes.

A variable that is constant in time cannot be vital.
Generally speaking, a measurement of a specific bodily function such as pulse rate, mental acuity,
blood serum level or forced expiratory volume is vital.
Likewise, any variable such as weight or body temperature, which cannot be measured without detecting whether the patient
is alive, is a vital variable.
Age is personal but not vital.
Other personal variables of a financial nature may not be vital,
and non-personal variables, even if they are health-related, are usually not vital.

Kalbfleisch and Prentice (2002, \S6.3) discuss various aspects of time-dependent processes
and their use as covariates in survival analysis.
Their definition of an {\it external variable} coincides with an exogenous variable as defined below;
their definition of an {\it internal variable} (not external) is not the same as a vital variable,
but the motivation is similar.
A variable may be neither vital nor external.

\section{Independent evolution}
Vitality is a property of a variable in isolation.
Independence is a property of a pair of variables or a set of variables.
For variables in the sense of temporal processes, independence is a very strong property
seldom satisfied by any pair in medical work.
For example, the death rate in certain cities is affected by the weather,
both extreme cold and extreme heat being contributing factors,
so survival time is not independent of the outside air temperature.
However, a weaker, and arguably more relevant, condition may be satisfied
such that the weather \emph{evolves independently} of the death rate.

Let $X, Y$ be two continuous-time stochastic processes,
and let $\H_t^{XY} = \H_t^X \otimes\H_t^Y$ be the $\sigma$-field generated by all past values $(X_s, Y_s)_{0\le s\le t}$.
The temporal evolution of $Y$ is governed by its transition probabilities,
or, more generally, by conditional probabilities of the type $\pr(Y \in A \given \H_t^Y)$
for various future events~$A$.
The temporal evolution of the pair is also governed by conditional probabilities
$\pr((Y, X)\in A\times B\given \H_t^{XY})$, and if we choose $B$ to be the entire $X$-space,
the conditional probability becomes $\pr(Y \in A \given \H_t^{XY})$.
By definition, the average value of $\pr(Y \in A \given \H_t^{XY})$, averaged over past $X$-values,
is equal to $\pr(Y \in A \given \H_t^{Y})$.
If, however, $\H_t^Y$ is sufficient to determine the conditional probabilities, i.e.,~if
\begin{equation}\label{iev}
\pr(Y \in A \given \H_t^{XY}) = \pr(Y \in A \given \H_t^Y),
\end{equation}
for each event~$A$,
then the subsequent trajectory of $Y$ is independent of past $X$-values.
In the symbolism of Dawid (1979),
\[
Y\indep \H_t^X \given \H_t^{Y}
\]
for every~$t$, and we say that \emph{$Y$~evolves independently of~$\H^X$}.

Independent evolution captures a certain asymmetric relation between two temporal processes.
Roughly speaking, the death rate $Y$ is not independent of the weather,~$X$,
but the weather evolves independently of the death rate $X\indep\H_t^Y \given \H_t^X$.
Conversely, the conditional hazard or death rate at time~$t$ given the entire weather trajectory
depends only on current and past values, not on future weather patterns,
so the death rate is driven by current and past weather patterns.
(This is a natural mathematical assumption, not a meteorological fact.)
In such cases, we say that the weather is \emph{statistically exogenous} for the response
(Robins, 1999),
or {\it external\/} in the sense of Kalbfleisch and Prentice (2002, p.~196).

In a similar manner with roles reversed,
the frequency of medical monitoring may depend explicitly on disease severity,
but the disease evolution may, in certain circumstances, be independent of the configuration of past monitoring times.
Conversely, the monitoring rate is driven by the disease history.
In the first example, the external variable evolves independently of disease and death;
in the second, the disease evolves independently of monitoring activity
(on the presumption that monitoring is passive or that the disease is incurable, and no
intervention is feasible).
The terms {\it optional sampling,} {\it sequential sampling,} and {\it optional stopping\/} are frequently used
in this setting (Dawid, 1979).

Most examples of external processes are also ancillary, in the sense that their
distribution is independent of the parameters of interest.
These two concepts are mathematically unrelated.
Independent evolution is a probabilistic property of each process individually.
Ancillarity is a distributional property of the {\it parameterized set\/} of processes in the model,
not a property of individual processes.

Independent evolution is a strong property of the joint distribution, but it is
not nearly so strong as complete independence.
Unlike independence, it is an asymmetric relationship;
if $Y$ evolves independently of $X$, then $X$~does not usually evolve independently of~$Y$.
On the other hand, if $Y$~evolves independently of~$X$,
and $X$~also evolves independently of~$Y$,
it is natural to ask whether the two processes are independent.
The answer is negative, but they are conditionally independent given the initial value
$(X_0, Y_0)$.

In the preceding discussion $\H_t^{XY}$~is the $\sigma$-field generated by
all values occurring at or before time~$t$.
More generally, the sequence $(\H_t^{XY})_{t\ge 0}$ is a filtration generated by some,
but not necessarily all, past values,
and condition (\ref{iev}) states that $Y$~is independent of $\H_t^X$ given $\H_t^Y$.
For example, $\H_t^{XY}$ could be the $\sigma$-field generated by the random variables $(X_s, Y_s)$
for integer times $s\le t$.
In that case, $Y \indep \H_t^X \given \H_t^Y$ is a statement not only about the
subsequent evolution of~$Y$, but also about values at earlier non-integer time points.
The same statement interpreted in reverse says that the conditional
probability of each event in $\H_t^X$ given $Y$ depends only on
those past values that are included in $\H_t^Y$.

\section{Joint models}
Let $X$ be a variable in the sense of a temporal stochastic process, and let $T$ be the survival time.
The term \emph{joint model} refers to the joint distribution of the pair $(X, T)$,
or equivalently, the pair $(X, \tilde T)$.
If $X$~is vital, the joint distribution is degenerate, and nothing further needs to be said.
Hence, without loss of generality, we assume that $X$ is non-vital.
The literature on joint models is very extensive, and no attempt is made here to review it.
For an overview, see
Henderson Diggle and Dobson (2000),
Tsiatis and Davidian (2004)
or Rizopoulos (2012).

The mathematical strategy most commonly employed for the construction of a joint process begins with
an unobservable, or latent, process $\eta$ such that $T \indep X \given\eta$.
For purposes of illustration, if $X$~is a real-valued process, we may choose $\eta$ to be a zero-mean Gaussian process
with covariance function $K$, followed by
\begin{eqnarray}
\nonumber
X(t) &=& \eta(t) + \epsilon(t) \\
\label{jm2}
-\log\pr(T > t \given \eta) &=& \int_0^t h(\eta(s)) \, ds ,
\end{eqnarray}
where $\epsilon(t)$ is white-noise measurement error,
and $h(\eta(t)) \ge 0$ is the conditional hazard function given~$\eta$.
This construction ensures that $\eta$ evolves independently of $(X,\tilde T)$,
and that $T$ and $X$ are not independent.

The conditional distribution of $\eta$ given~$X$ is Gaussian
\[
\eta \given X \sim N(K(I+K)^{-1} X,\;  K(I+K)^{-1})
\]
with conditional mean linear in~$X$.
The conditional survivor function given $X$
\begin{eqnarray}\label{gaussianintegral1}
\pr(T > t \given X) &=& E\biggl( \exp\Bigl(-\int_0^t h(\eta(s))\, dx \Bigr) \Given X \biggr)
\end{eqnarray}
reduces to a Gaussian integral, albeit infinite-dimensional.
For any given covariance function, it is easy to check numerically
whether the conditions for independent evolution of $X$ and $\tilde T$ are satisfied,
for example, by checking whether the conditional survivor function (\ref{gaussianintegral})
is or is not independent of future $X$-values.
It is possible that the independent-evolution condition may be satisfied by
certain special covariance functions or by special choice of parameters in (\ref{jm2}),
but no such parameters are known apart from degeneracies such as $\epsilon(t) = 0$ for all~$t$.
In general, it appears that the conditional survivor function
depends not only on past $X$-values, but also on future $X$-values.
In other words, the process~$X$ is neither vital nor statistically exogenous.

In the literature on joint models, 
the latent process is commonly referred to as the patient's \lq true state of health\rq.
In fact, (\ref{jm2}) implies that $\eta$ satisfies neither of the conditions for vitality,
so the \lq true state of health\rq\  at time~$t$ is not sufficient to determine
the most basic vital fact---whether or not the patient is alive.
Among the eight variables $X$ listed as examples in section~1,
five are naturally regarded as vital.
A further two are presumed to satisfy the independent-evolution condition,
one example for $X\indep \H_t^Y \given \H_t^X$ and one for $Y\indep\H_t^X \given \H_t^Y$,
with $Y=\tilde T$.
That leaves only one candidate---the value of the patient's retirement portfolio---as a
plausible example of a variable that satisfies neither condition.

If $Y$ were a time-evolving risk factor such as the measured ozone pollution level,
a reasonable case might be made that $\eta(\cdot)$ is the \lq true pollution level\rq.
That usage is in keeping with Besag and Higdon (1999).
Its justification rests on the presumption that the true pollution level varies
in time as prescribed by~$K$, usually continuously,
 so that any white-noise component in $Y$ must be associated with pure measurement error.
Even in that situation, it is not easy to rationalize the dependence
of the conditional survivor function $\pr(T > t \given Y)$ on future measured
pollution levels unless each cadaver contributes subsequently to pollution or to measurement error.

\section{Stochastic specification}
Let $Y$ be a vital variable, i.e.,~a stochastic process in continuous time with state space~$\State$,
and let $T > 0$ be the survival time.
It is always possible to recode $T$ as a Boolean process and to include it as a component in~$Y$,
so there is no loss of generality in the restriction to vital variables.

In order to specify the joint distribution of $Y$, it is necessary and sufficient to specify,
for each finite collection of ordered time points $\bft=(t_1< \cdots < t_k)$,
the joint distribution $P_\bft(\cdot)$ of the values $(Y_{t_1},\ldots, Y_{t_k})$ in $\State^k$.
These joint distributions are subject to the standard Kolmogorov consistency conditions for a stochastic process.
The density function relative to a suitable product measure in $\State^k$ is denoted by $p_\bft(y)$.

Although these finite-dimensional distributions are sufficient to determine the joint distribution
of the process, it is sometimes helpful to specify the same distribution in an alternative manner.
We may ask for the joint distribution $Q_\bft(\cdot, dt)$ of the values $(Y_{t_1},\ldots, Y_{t_k})$
together with the survival time $T>0$, in $\State^k \times \Real^+$.
These distributions are also subject to Kolmogorov consistency conditions.
In addition
\[
P_\bft(A) = Q_\bft(A\times\Real^+) = \int_0^\infty Q_\bft(A, dt),
\]
for arbitrary events $A\subset\State^k$, so the mapping from $Q$ to $P$
is a one-dimensional integral over survival times.

The joint distribution $Q_\bft$ also determines the clinical predictive distribution
\[
\pr(T \in dt \given (y, \bft)) = q_{\bft}(y, dt) / p_\bft(y)
\]
of the survival time given the finite sequence of values $y=(y_1,\ldots, y_k)$
occurring at earlier appointment times $t_1,\ldots, t_k$.
Ordinarily, the patient is alive at time $t_k$, so the conditional distribution
is supported on $(t_k, \infty)$.
The clinical predictive distribution is not to be confused with the conditional distribution
given the past history up to time~$t_k = \max(\bft)$
because the latter is a function of the entire trajectory, which is seldom observed
in a clinical setting.

\section{Sampling distributions}
A point process on the real line may be sampled by counting events in a fixed domain,
or by measuring inter-arrival times for a fixed number of events.
The two sampling distributions are of a different nature on different spaces, but they are mutually consistent,
equivalent, and they are both determined by the point process.
Likewise, in a survival study, the values may be acquired in more than one way,
and it is necessary to describe the sampling protocol before the sample space and the
sampling distribution can be specified.
In particular, the processes are defined continuously in time, but seldom observed continuously.
In addition, values of non-vital variables may be recorded post mortem, but this is rarely done
because such values would seldom be considered relevant to the objectives of the study.

We consider first the simplest data-acquisition scheme in which measurements are made on an
arbitrary schedule, specified at the time of recruitment, for a fixed period~$L$ comprising $k$ measurement times,
which are not necessarily the same for each patient.
If, however, the patient dies before the end of the observation period, the time of death is also recorded.
The sample space for this scheme is the disjoint union $\State^k \cup (\State^k\times(0,L))$.
The probability density is $p_\bft(y)$ at $y$ in~$\State^k$ plus
$q_\bft(y, t)$ on the product space $\State^k\times (0,L)$.

The presumption here---that measurement continues post mortem---is not necessarily realistic,
but the implications are worth pursuing.
First, if $Y$ is a purely vital process with a single absorbing state,
the post-mortem value is fixed and uninformative,
so it is immaterial whether the value is explicitly recorded.
Second, if $Y$~is purely vital with multiple absorbing states,
the post-mortem value is a random constant, which could be informative for cause of death.
A typical health trajectory prior to accidental death might be
quite different than the a typical trajectory prior to death from leukaemia.

The more interesting case is one in which $Y = (Y_0, Y_1)$, where $Y_0$ is purely vital and $Y_1$ is non-vital.
For example $Y_1$ might be a time-evolving risk factor.
If $Y_1$ evolves independently of $Y_0$ given the history of both, the
post-mortem evolution of $Y_1$ is independent of the vital history,
and it can reasonably be argued on that contextual basis that the
post-mortem evolution is not relevant for patient health.
On the other hand, if $Y_1$ does not evolve independently of $Y_0$,
this argument no longer applies, and the subsequent values do affect the likelihood.
For example, the irrelevance argument does not apply to joint models using the construction in section~4.

For the second data-acquisition scheme, the sampling times
$\bft = (t_0 < t_1 < \cdots)$ are chosen randomly with $t_0=0$
in such a way that $Y$ evolves independently of $X$ given the \emph{observed} history.
Here, $X(t_j) = t_{j+1}$ is the sequence of sampling times recorded as a c\`adl\`ag step function.
The first observed value is a pair $(Y(t_0), t_1)$,
the next value $(Y(t_1), t_2)$ occurs at time $t_1 > 0$,
followed by $(Y(t_2), t_3)$ at time $t_2 > t_1$, and so on,
so $\H_t^{XY}$~is the $\sigma$-field generated by the variables
$(Y(t_j), t_{j+1})$ for $t_j \le t$.
Sampling terminates at a fixed time $L$ or at death, whichever comes first.
If death occurs, the time is recorded.
This procedure gives rise to a sequence of times $\bft$ of random length,
and a sequence of values $y\in \State^{\#\bft-1}$, one value for each sampling time except for the last.
The independent-evolution condition implies
\[
\pr(Y \in A \given \H_{t}^{XY}) = \pr(Y \in A \given \H_{t}^Y)
\]
for every event $A$.
This is equivalent to the condition that the distribution of $t_{j+1}$ given $Y$ is a function
of previous observed values only, i.e.,~a function of $t_0,\ldots, t_j$ and $Y_{t_0},\ldots, Y_{t_j}$,
which is the sequential conditional independence condition of
Dempsey and McCullagh (2016).

A consequence of the independent-evolution condition is that the joint density of the times and values is either
\begin{equation}\label{censored}
p_\bft(y) \times \prod_{j\ge 0} p(t_{j+1} \given \H_{t_j}^{XY})
\end{equation}
if no failure occurs, or
\begin{equation}\label{uncensored}
q_\bft(y, t) \times \prod_{j\ge 0} p(t_{j+1} \given \H_{t_j}^{XY})
\end{equation}
if failure occurs at time~$t \le L$.
In this setting, $p(s \given \H_{t}^{XY})$ is the conditional density at $s > t$ of the
next scheduled appointment given the observed values up to time~$t$.
Note that $\#\bft$ is a random variable whose distribution is not independent of~$Y$.

An important aspect of the preceding derivation is that the next appointment time
is scheduled and recorded at the previous appointment, so every off-schedule appointment
is detectable, and is a breach of protocol.

\section{Likelihood function}
Consider now a family of probability distributions indexed by $\theta\in\Theta$ for a vital process~$Y$,
which is sampled according to one of the schemes described in the preceding section.
The joint density of the $Y$-values at a fixed configuration $\bft$ of sampling occasions is
$p_\bft(y; \theta)$.
The joint density of values and survival time is $q_\bft(y, t; \theta)$,
so that $p_\bft(y; \theta) = \int_0^\infty q_\bft(y, t; \theta)\, dt$.

The second factor in (\ref{censored}) is a probability distribution determined
entirely by the experimental protocol.
In other words, the second factor is parameter-free
and does not contribute to the likelihood function.
Although the sample space and the sampling distribution depend heavily on the sampling scheme,
the likelihood function for the second data-acquisition scheme is the same as if
the sampling times were fixed in advance.
The independent-evolution assumption is essential for this conclusion, so all off-schedule appointments
should be noted and treated with caution.
For an instance of this, see Liest\o l and Anderson (2002).

On the assumption that processes for distinct patients are independent,
the likelihood function is the product over patients, each censored record contributing a factor
$p_\bft(y; \theta)$,
each uncensored record contributing $q_\bft(y, t; \theta)$.
Of course, the values $(\bft, y)$ or $(\bft, y, t)$ are patient-specific,
and the sequence lengths $\#\bft$ also vary from one patient to another, even in
a setting where the responses for distinct patients are identically distributed.

In the presence of covariates such as age or treatment,
the response distributions for two patients having different $x$-values may be different.
Denote the response density at $y$ and $(y,t)$ for one patient with covariate $x$
by $p_\bft(y; x, \theta)$ and $q_\bft(y, t; x,\theta)$ respectively.
Then the likelihood function function is a product of these factors,
one density term for each patient.

\section{A Gaussian survival process}
\subsection{Background}
Up to this point, the emphasis has been on general principles,
no attempt being made to construct specific survival processes for use in applications.
We consider now an example of a survival process of the simplest type, one in which $Y$ is a purely vital variable
with state space
$\State = \Real \cup \{\flat\}$;
either $Y_i(t)$ is a real number, in which case patient~$i$ is alive at time~$t$,
or $Y_i(t) = \flat$ in which case the patient is dead at time~$t$.
In this setting, where post-mortem values are fixed and non-random,
it is immaterial which of the two sampling schemes described in section~6 is used.
For the second scheme, the assumption that $Y$ evolve independently of the scheduled appointment times is crucial.
Initially, for simplicity, it is assumed that there are no covariates,
so the observations for distinct patients are independent with the same distribution.

We first derive the joint density function $q_\bft(y, t)$ for the values and the survival time.
Let the survival density be~$f$.
Let the conditional distribution of $Y$ given $T=t$ be Gaussian on the interval $[0, t)$
with conditional moments
\begin{eqnarray}
\nonumber
E(Y_s \given T=t) &=& \mu_t(s) \cr
\label{gcd}
\cov(Y_s, Y_{s'} \given T=t) &=& K_t(s, s')
\end{eqnarray}
for $0 \le s, s' < t$.
The conditional covariance function is necessarily positive definite,
but the specification is otherwise unrestricted.
On the assumption that $T < \infty$ with probability one,
Dempsey and McCullagh (2016) use a family of conditional mean functions
\begin{equation}\label{additive}
\mu_t(s) = \alpha(t) + m(t-s),
\end{equation}
which is an additive function of the survival time $t$ and the revival time $t-s$.
Here, $m(\cdot)$~is the \emph{characteristic mean curve} of the process in revival time.
Additivity on this scale was found to be much more effective than additivity in $t$ and ~$s$.

In practice, the conditional covariance function must include
an additive random constant for each patient,
a non-trivial temporal process for each patient
and a white noise term independent for each patient and each time.
The first two should be independent and identically distributed for each patient.
Other covariance terms may also be needed, depending on the context.

If the design includes a treatment effect, the survival process for the active
treatment levels may be different in distribution from the the control process.
To accommodate the effect of treatment on the health process, the additive model may be modified
so that the conditional mean given $T=t$ for a patient with treatment level~$x$ is
\[
\mu_t(s) = \alpha(t) + m_{x}(t-s).
\]
Additivity is retained, but the characteristic mean curve depends on the treatment arm,
the simplest version being $m_x(t-s) = m_0(t-s) + \beta_x$, so that the mean curves are parallel
in revival time.
Given two patients, one surviving for three years with $x_1=1$ and one for five years with $x_2=2$,
the conditional means at time $t_i-z$, i.e.,~$z$~years prior to failure are
\begin{eqnarray*}
\mu_3(3-z) &=& m_0(3) + m_1(z), \\
\mu_5(5-z) &=& m_0(5) + m_2(z)
\end{eqnarray*}
respectively.
Note that $\mu_3(s)$ is not defined for $s \ge 3$, so it is not possible to compare 
conditional means at arbitrary fixed times,
but it is possible to compare them for arbitrary fixed revival time $t-s$, as indicated here.
The difference $\mu_5(5-z) - \mu_3(3-z)$ is a sum of two parts,
one related to the effect of the differing survival times,
and the second, $m_1(z) - m_2(z)$, associated with the effect of treatment
on the patients' health while they are alive.
The function $\mu_t(t-z)$ is defined for all $t, z > 0$, but only $z\le t$ is typically needed.

It is important to bear in mind that
the first component $Y_i(0)$ is measured at recruitment,
and that, modulo covariate information, all patients are on an equal footing at this point.
More generally, $x_i = x_j$ implies that $Y_i(0) \sim Y_j(0)$ have the same distribution regardless of
the treatment arm to which they are subsequently assigned. 
Thus, treatment is a time-varying function whose value $x_i(s)$ is constant for $s > 0$,
but $x_i(0) = x_j(0)$ is the same null level for every patient.\strut

In the following description, $\bft\subset[0,\infty)$ is an arbitrary ordered set of $k$ sampling times,
and $y\in\Real^k$ is a real vector.
The joint density at $(y, t)$ of the values and the survival time is
\begin{equation}\label{gsdq}
q_\bft(y, t) = f(t) \times(2\pi)^{-k/2} |\Sigma|^{-1/2} \exp\Bigl(- (y-\gamma)'\Sigma^{-1} (y-\gamma)/2\Bigr),
\end{equation} 
for $t > \max(\bft)$, and zero otherwise.
Here, $\gamma = \mu_t[\bft]$ is the conditional mean function
and $\Sigma=K_t[\bft,\bft]$ is the conditional covariance function,
both evaluated at the sampling times.

The joint marginal density at $y$ is obtained by a one-dimensional integral
\begin{equation}\label{gsdp}
p_\bft(y) = \int_0^\infty q_\bft(y, t)\, dt,
\end{equation}
which may be restricted to the range $t > \max(\bft)$. 
The dependence of the conditional moments on~$t$, means that this integral must be computed numerically.
The marginal density at any other point in $\State^k$,
for example an interval-censored record with one or more trailing components of $y$ equal to $\flat$,
is obtained by integration of (\ref{gsdq}) over a finite range.

The joint density function for all patients is a product of $n$ factors,
one factor of type (\ref{gsdq}) for each uncensored record,
and one factor of type (\ref{gsdp}) for each right-censored record.

The state of affairs is only slightly more complicated if the process has more than one absorbing state,
which is also recorded at the time of death.
In that case, $f$ is the joint distribution of time of death and the absorbing state,
while $q_t(s)$ is the conditional distribution of $Y$ given the failure time and absorbing state.
The situation is considerably more complicated if $Y$ has multiple components,
including non-vital components such as a time-evolving risk factor.

\subsection{Parameter estimation}
Suppose that the parameter vector can be partitioned into two components $\theta = (\lambda, \psi)$,
the first related to the distribution of survival times,
and the second related to the conditional distribution given the survival time.
Both components contain parameters related to covariate and treatment effects.

Let $[n]=\{1,\ldots, n\}$ be the set of records, and $C \subset[n]$ the subset of censored records.
For an uncensored record, $t$~is the survival time;
for a censored record, $t$~is the censoring time.
The joint density is a product of four factors:
\[
\prod_{i\in \bar C} q_\bft(y_i, t_i;\theta) \prod_{i\in C} p_{\bft}(y_i;\theta) = 
\prod_{i\in \bar C} f(t_i; \lambda) \prod_{i\in C} S(t_i;\lambda) \times
\prod_{i\in \bar C} \frac{q_\bft(y_i, t_i;\theta)} {f(t_i;\lambda)}
\prod_{i\in C} \frac{p_\bft(y_i;\theta)} {S(t_i;\lambda)}
\]
where $f(t; \lambda)$ is the density function and $S(t; \lambda)$ is the survivor function.
The marginal likelihood based on the observed survival and censoring times only,
is the product of the first two factors, which
is a function of $\lambda$ alone.

The third factor is the conditional distribution of $Y$ given the failure times for those
patients who have been observed to fail.
The ratio
\[
\prod_{i\in\bar C} \frac{q_\bft(y_i, t_i;\theta)} {f(t_i;\lambda)},
\]
is a standard Gaussian likelihood function depending only on~$\psi$.
The final term for censored records is a function of both parameters.

This factorization is extremely convenient because it allows us to obtain
consistent estimates of both parameters in a simple manner using standard software
for survival analysis and standard software for Gaussian models,
and to check various aspects of the model formulation.
In addition, by examining the fourth factor for censored records alone, it is possible, and strongly advised,
to check whether the behaviour of the health process for censored records is
compatible with that for uncensored records.

After these preliminary calculations have identified a set of candidate models for comparison,
the joint likelihood function may be computed and maximized.
This last step, and only this step, requires specialized software for computing the
one-dimensional integral in~(\ref{gsdp}).

Computational matters are beyond the scope of this paper, but details are provided in Dempsey and McCullagh (2016).

\section{Time-evolving exposure}
\subsection{Statistically exogenous variable}
Let $(X, Y)$ be a bivariate temporal process in which $Y$ is purely vital,
and $X$ is an exogenous exposure variable.
As the term suggests, a vital variable is a health-related response;
an exogenous variable may be indirectly health-related, such as an environmental exposure or risk factor.
This terminology implies that the focus of attention is on the behaviour of $Y$
in response to~$X$, suggesting that $X$ drives $Y$ rather than vice-versa.

Let $\H_t^{XY} = \H_t^X\otimes\H_t^Y$ be the complete history of the joint process up to time~$t$.
We say that $X$~is \emph{statistically exogenous for patient health} if
the independent-evolution condition
\[
X \indep \H_t^Y \given \H_t^X
\]
is satisfied for every~$t$.
An equivalent statement is that the conditional distribution of health values up to time~$t$
given the entire trajectory of the exposure factor depends only on past exposures.
Roughly speaking, future exposure has no effect on current or past health.

For the remainder of this section, $X$~is statistically exogenous for~$Y$.

\subsection{A special Gaussian process}
One way to construct a joint survival process is to factor the joint density
in the obvious way:
\[
p(X)\cdot p(T \given X) \cdot p(Y \given X, T).
\]
The first part is the marginal distribution of the exogenous process;
the second is the distribution of the survival time given the {\it entire\/} exposure trajectory,
and the third is the conditional distribution given $X, T$.
Since $X$ is statistically exogenous by assumption, the conditional survivor function
$\pr(T > t\given X)$ is a function of the past history only,
and the third factor has a similar property.

Suppose first that $Y = \tilde T$, so the third factor is degenerate.
We ask only for a joint distribution  for $(X, T)$ in the sense of section~4,
but with the additional requirement that $X$ be exogenous.
For definiteness let $X\sim\GP(\mu, K_0)$ be a continuous-time Gaussian process with mean $\mu$
and covariance function~$K_0$.
This process continues indefinitely in time, and may even be stationary.
The second factor is the conditional survival distribution, or the conditional hazard density at time~$t$ given $X$,
which, for simplicity, we take to be $h(X(t))\, dt$,
depending only on the current exposure.

The preceding paragraph specifies the process but does not directly yield finite-dimensional
distributions of the type needed for likelihood calculations in statistical work,
where the exposure process is not observed continuously in time.
For example, the conditional distribution given $X$ is not to be confused with the
conditional distribution given the {\it recorded\/} values of~$X$.
For a finite collection of $k$~time points $\bft$, the joint density $q_\bft(x, t)$ of the exposure values
and the survival time at $(x, t)$ is the product of two factors.
The first is the Gaussian density at $x\in\Real^k$ with mean $\mu[\bft]$
and covariance $K_0[\bft,\bft]$, both evaluated at the indicated sampling points.
The second is the conditional survival density given $X[\bft]=x$, which is a Gaussian integral
\begin{equation}\label{gaussianintegral}
f(t \given X[\bft]) = E\biggl(h(X(t)) \exp\Bigl(-\int_0^t h(X(s))\, ds\Bigr) \Given X[\bft] \biggr),
\end{equation}
similar to that arising in (\ref{gaussianintegral1}) for a joint model.
This is the contribution to the log likelihood function for an uncensored record.
In the case of a record censored at $t$, the conditional density (\ref{gaussianintegral}) is
replaced by the conditional survivor function, which requires an additional one-dimensional
integral over survival times.

It is quite possible in this setting that $t < \max(\bft)$, so the exposure process is measured post mortem,
or possibly also pre-recruitment.
In a medical sense, individual health cannot be affected by post-mortem exposure,
but the integral (\ref{gaussianintegral}) is affected, and the likelihood function also.
In this respect, an external exposure process is not equivalent to a time-varying covariate
such as age or sex-age interaction,
which is known as a function of time.

These are the calculations needed to study the effect of a time-evolving exposure factor on
the distribution of survival times.
Where additional health measurements are involved, the likelihood function has a third factor.
Suppose that $Y$ is a Gaussian health process of the type described in section~8,
with conditional moments
\begin{eqnarray*}
E(Y_s \given X, T=t) &=& \mu(s, t, X(s)) \\ 
\cov(Y_s, Y_{s'} \given X, T=t) &=& K_t(s, s')
\end{eqnarray*}
for $s, s' < t$.
At this point, we have made the standard linear-model assumption that the variance is
constant and independent of treatment and exposure variables.
On account of the dependence of $E(Y_s)$ on the contemporaneous exposure $X(s)$ only,
the third factor is a Gaussian density of the type discussed in section~8, which requires
no complicated integration.
Otherwise, if $E(Y_s\given X, t)$ were a non-trivial function of past exposures at
times other than those in~$\bft$,
it would be necessary to compute a Gaussian conditional expected value as in (\ref{gaussianintegral}).

For a record censored at~$t$, it is necessary to integrate the triple product over
survival times, which is a one-dimensional integral.
For this model at least, the additional computations required to accommodate a sequence
of health values is relatively modest.

\def\journal#1{{\it #1}}
\def\title#1{{\it #1}}
\def\vol#1{{\bf#1}}
\section{References}
\everypar={\parskip 0.5pc \parindent 0pt \hangindent 10pt }
\noindent
Besag, J. and Higdon, D. (1999). Bayesian analysis of agricultural field experiments (with
discussion). J. Roy. Statist. Soc. Ser. B 61 691–746.


Dawid, A.P. (1979)
Conditional independence in statistical theory (with discussion).
{\it J.~Roy.\ Statist.\ Soc.}~B 41, 1--31.

Dempsey, W. and McCullagh, P. (2016)
Survival models and health sequences.
AeXiv1301.2699


Henderson, R., Diggle, P. and Dobson, A. (2000)
Joint modelling of longitudinal measurements and event time data.
\journal{Biostatistics} \vol1, 465--480.

Kalbfleisch, J. and Prentice, R. (2002)
{\it The Statistical Analysis of Failure Time Data}.
2nd~ed. New York: John Wiley \& Sons.


Liest\o l, K, and Andersen, P.K. (2002)
Updating of covariates and choice of time origin in survival analysis:
problems with vaguely defined disease states.
\journal{Statistics in Medicine} \vol{21}, 3701--3714.

Rizopoulos, D. (2012)
{\it Joint Models for Longitudinal and Time-to-Event Data.}
Chapman and Hall.

Robins, J.M. (1999)
Marginal structural models versus structural nested models as tools for causal inference.
In {\it Statistical Models in Epidemiology, the Environment and Clinical Trials}
(editors, E.~Halloran and D.~Berry), 95--134.
New York: Springer-Verlag.

Tsiatis, A.A, and Davidian, M. (2004)
Joint modeling of longitudinal and time-to-event data: an overview.
\journal{Statistica Sinica} \vol{14}, 809--834.

\end{document}